\documentclass[]{aipproc}

\layoutstyle{8x11single}

\SetInternalRegister\hbadness{8000} 

\newcommand\doingARLO[2][]{%
  \ifx\mmref\undefined #1\else #2\fi
}
  
\newcommand{ \be }{\begin{equation}}
\newcommand{ \ee }{\end{equation}}
\newcommand{ \bea }{\begin{eqnarray}}
\newcommand{ \eea }{\end{eqnarray}}
\newcommand{ \la }{\langle}
\newcommand{ \ra }{\rangle}

\newcommand{ \mpt } {{\la p_t \ra }}
\newcommand{ \sigmampt } { \sigma_\mpt }
\newcommand{ \impt } {{\overline {p_t} }}
\newcommand{ \sigmapt } { \sigma_{p_t,inclusive} }
\newcommand{ \mmpt } { \la \la p_t \ra \ra}

\newcommand{ \sigmadyn } {\sigma_{\mpt ,dynam} }
\newcommand{ \sigmastat } {\sigma_{\mpt ,stat} }
\newcommand{ \phisubpt } {\Phi_{p_t} }

\begin{document}

\title {
Multiplicity and mean transverse momentum fluctuations in Au + Au
 collisions at RHIC.
}


\author{Sergei A. Voloshin$^*$ for the STAR Collaboration}
{
  address={$^*$Wayne State University, 666 W. Hancock, Detroit,
Michigan},
  email={voloshin@physics.wayne.edu}
}

\copyrightyear  {2001}

\begin{abstract}
Strong modification of event-by-event fluctuations in global
observables was proposed by several authors as an indicator 
of the phase transition in system evolution. 
We report the results of the analysis of STAR TPC data on 
the charged multiplicity  and mean transverse momentum
fluctuations in Au+Au collisions at $\sqrt{s_{NN}}$=130 GeV. 
We discuss the centrality dependence of the observed fluctuations 
and compare the results with available data from lower collision 
energies and ISR data. 
\end{abstract}

\date{\today}

\maketitle

\section{Introduction}

The question of the hadronization of the system created in
ultrarelativistic heavy ion collisions is one of the most interesting
questions in this field. 
Does the system evolution include the phase transition? 
Event-by-event fluctuations of the mean transverse momentum and/or
multiplicity, are considered to be one of the important tools 
to identify such a phase transition.
The fluctuations depend on the nature of the phase transition. 
A first order phase transition may lead to 
large fluctuations in energy density
due to formation of QGP droplets{\cite{vanHove,kapusta}}. 
Second order phase transitions may lead to divergence in the specific heat; 
it would also increase the fluctuations in energy density due to
long range correlations in the system~\cite{stephanov98,stephanov99}.
One could observe them as fluctuations in mean transverse momentum
if matter freezes out at the critical temperature
$T_c$~\cite{stodolsky,shuryak98}.
The system freeze-out near the QCD tri-critical and critical points 
would also lead to change in the fluctuations 
pattern~\cite{stephanov98,stephanov99,berdnikov}.  
There exist also
other sources of the abnormal dynamical fluctuations.
The formation of the disoriented chiral condensate~\cite{dcc} 
should lead to large fluctuations in the ratio 
of neutral to charged pion yields at low $p_t$, 
and consequently to fluctuations in $\mpt$ of charged particles.  
Jets may give rise to event-by-event fluctuations
in the high $p_t$ region. 

For the multiplicity fluctuations, predictions range from enhanced multiplicity
fluctuations connected to the production of QGP droplets 
and nucleation processes in a first order 
phase transition, to a strong suppression of fluctuations as a consequence
of rapid freeze-out just after the phase transition. 
In this case, final state values of
conserved quantities, such as net electric charge, baryon number, and
strangeness would not be strongly modified from their value in the
QGP stage. The fluctuation of these quantities 
could be reduced by a factor 2 -- 4 if a QGP is
produced ~\cite{Asakawa00,Jeon00}. 
The production and size of QGP droplets may critically depend 
on the collision impact parameter,
leading to centrality dependence of the fluctuations to be an
important observable. If the fluctuations are due to the
particle production via any kind of clusters (e.g., resonances, strings
(mini)jets, independent $NN$-collisions, etc.) and the relative
production of such clusters do not change with centrality,
the correlation strength should be inversely proportional to the particle
multiplicity. New physics should appear as a deviation
from such a dependence.   

Here, we report on our measurement of charged particle multiplicity 
and mean $p_t$ fluctuations 
as a function of collision centrality in Au + Au
collisions at an energy of $\sqrt{s_{NN}}=130$ GeV. 
Specifically, for the case of multiplicity fluctuations 
we discuss fluctuations in
the difference of the relative number of produced positive and negative charged
particles measured in a fixed rapidity range:
\be
\nu_{ab} = \la \left(\frac{N_a}{\la N_a\ra}
-\frac{N_b}{\la N_b\ra}\right)^2 \ra,
\ee
where $N_a$ and $N_b$ are the corresponding multiplicities. 
One can consider fluctuations in the production of charged
particles, $N_+$ and $N_-$, as well as in the specific cases
of proton and anti-proton, $N_p$ and $N_{\overline{p}}$, and positive 
and negative kaons, $N_{K+}$ and $N_{K-}$. 
The former amounts to a measurement of net electrical charge fluctuations, 
whereas the latter cases are related to net baryon number and net 
strangeness fluctuations, respectively. 
The quantity $\nu_{ab}$ used in our analysis
can be related to the fluctuations in multiplicity 
ratios~\cite{heiselberg,Jeon99}, but have the advantage of
not being limited to large particle multiplicities.

\subsection{Statistical and Dynamical fluctuations}

The fluctuations of any observable measured on an event-by-event 
basis consists of two parts: {\it statistical fluctuations}, 
the uncertainty in the measurement of a given quantity due 
to finite event multiplicity,
and {\it dynamical fluctuations} due to correlated particle production,
the subject of interest. 
The statistical fluctuations are the ones one would observe 
in the case of 
independent particle production provided that the single particle
inclusive quantities remain the same as in real events. 

{\bf Multiplicity fluctuations.}
A difficulty inherent in the interpretation of multiplicity
fluctuations is the elimination of effects associated with
uncertainties in the collision centrality, often referred to as volume 
fluctuations. They, in particular, induce a
positive correlation in particle production which does not depend on the
intrinsic dynamical properties of the colliding system. 
Fluctuations in the difference of relative multiplicities, $\nu_{ab}$, 
defined in Eq.~1, are however free from this problem. 
This analysis is thus restricted to the study of such relative
multiplicities. 

The magnitude of the variance, $\nu_{ab}$, is determined by both
statistical and dynamical fluctuations.  
Taking into account the fact that for independent particle
production the multiplicity fluctuations follow the Poisson distribution,
the dynamical fluctuations can be evaluated as
\be
\nu_{ab,dyn} = \nu_{ab} - \nu_{ab,stat}, \;\;\; 
\nu_{ab,stat} = \frac{1}{\la N_a \ra }+\frac{1}{\la N_b \ra}.
\ee
The value of $\nu_{ab,dyn}$ is determined by intrinsic
correlations between produced particles averaged over 
the pseudo-rapidity and transverse momentum range under study.
It is, to first order, independent of 
the detector acceptance (provided it is smaller than typical
long-range correlation scale of 1 -- 2 units of rapidity)
and particle detection efficiency
as can be seen from its relation to the inclusive two 
and single particle densities: 
\be
\nu_{ab,dyn}=R_{aa}+R_{bb}-2R_{ab},
\label{nudyn}
\ee
where  $R_{ab}$ is a correlation function similar to that
 used in multi-particle
production analysis~\cite{Foa75,Whitmore76,Boggild74} and defined as
\be
R_{ab}=\frac{\int_{\Delta Y} \rho_{2,ab}(\eta_a,\eta_b) d\eta_a d\eta_b}
   {\int_{\Delta Y} \rho_{1,a}(\eta_a) d\eta_a
     \int_{\Delta Y} \rho_{1,b}(\eta_b) d\eta_b  } -1,
\label{Rab}
\ee
where  $\rho_1(\eta)=dn/d\eta$, and  
$\rho_{2}(\eta_a,\eta_b)=d^2n/d\eta_a d\eta_b$
are single and two particle (pseudo)rapidity densities, respectively.

{\bf Mean transverse momentum fluctuations.}
We use the following notation for the mean transverse momentum 
in each event and average over all events in
a given event sample (centrality bin):  
\begin{equation}
\mpt = \frac{1}{M}{\sum_{i=1}^{M} p_{t_i}}, \;\;\;\;\;
\mmpt = \frac{1}{N_{evt}}{\sum_{j=1}^{N_{evt}}\mpt_{j}}.
\label{empt}
\end{equation}
The fluctuations in $\mpt$ becomes
\bea
\sigma^2_{\mpt}&=& \la \mpt^2 \ra -\mmpt^2 
= \frac{1}{M^2} \la \la \sum_i p^2_{t,i}+\sum_{i \neq j} p_{t,i}
p_{t,j} \ra \ra - \mmpt^2 
\\
&=&\sigmastat^2 +\frac{M-1}{M} \sigmadyn^2,
\;\;\;\;\;\; {\rm with} \;\;\;
\sigmastat^2=\frac{\la\la p_t^2\ra\ra
-\mmpt^2}{M}=\frac{\sigmapt^2}{M}.
\label{mptfluc}
\eea 
where $M$ is the multiplicity.
In Eq.~\ref{mptfluc} we have introduced the notation
for the non-statistical (dynamical) fluctuations in $\mpt$ as: 
\be
\sigmadyn^2 \equiv  \la \la p_{t,i} \, p_{t,j} \ra_{i\ne j} \ra 
                    - \la \la p_{t} \ra \ra^2,
\label{sigmadyn}
\ee
which is zero in the case of no correlation between particle 
transverse momenta.
As above,  in this equation the double angle brackets represent 
the average over particle pairs in a specified (pseudo)rapidity 
and transverse momentum window, and the   
average over all events with a given multiplicity.

\section{Results}

{\bf STAR detector and data selection.}
The STAR detector consists of many detector subsystems 
in a large solenoidal magnet with the Time Projection Chamber (TPC)
as the main tracking device.
Details about the STAR TPC can be found 
in reference{\cite{Thomas99}}.
In the year 2000, the TPC was operated in a 0.25~T magnetic field 
allowing tracking of particles with $p_t$ greater than 75~MeV/c. 
The experimental setup includes also the scintillator trigger 
barrel (CTB) surrounding the TPC and two Zero Degree Calorimeters (ZDCs) 
located at 18~m from  the center of the TPC.  
The ZDCs measure mainly fragmentation neutrons and are used 
in coincidence as the experimental trigger for
minimum bias events. 

We determine the centrality of the collision by the total 
reconstructed charged track multiplicity, $N_{ch}$, in the pseudorapidity 
region  $|\eta|<0.75$. 
Eight centrality bins used in this analysis are the same as
used in~\cite{flow} and correspond to
the fraction of the total geometrical  cross
section of the top 6\%, 11\%, 18\%, 26\%, 34\%, 45\%, 58\%, 
and 85\%, respectively.

A total of about 140 thousand minimum bias events with a primary vertex 
position within 75~cm of the center of the TPC have 
been used in this analysis.
The tracks are required to pass within 3~cm of the primary vertex 
and the ratio of the number of points on a track to the
maximum possible number of points is required to be  greater than $0.5$ 
in order to avoid double counting of tracks due to track splitting.
In this analysis we use only tracks within the pseudo-rapidity range 
$-0.75<\eta<0.75$ and transverse momentum 
$0.1<p_t<2.0$~GeV/c, where the track reconstruction efficiency is estimated
to be  $90 \pm 10\%$.

\begin{figure}
{\includegraphics[height=.25\textheight]{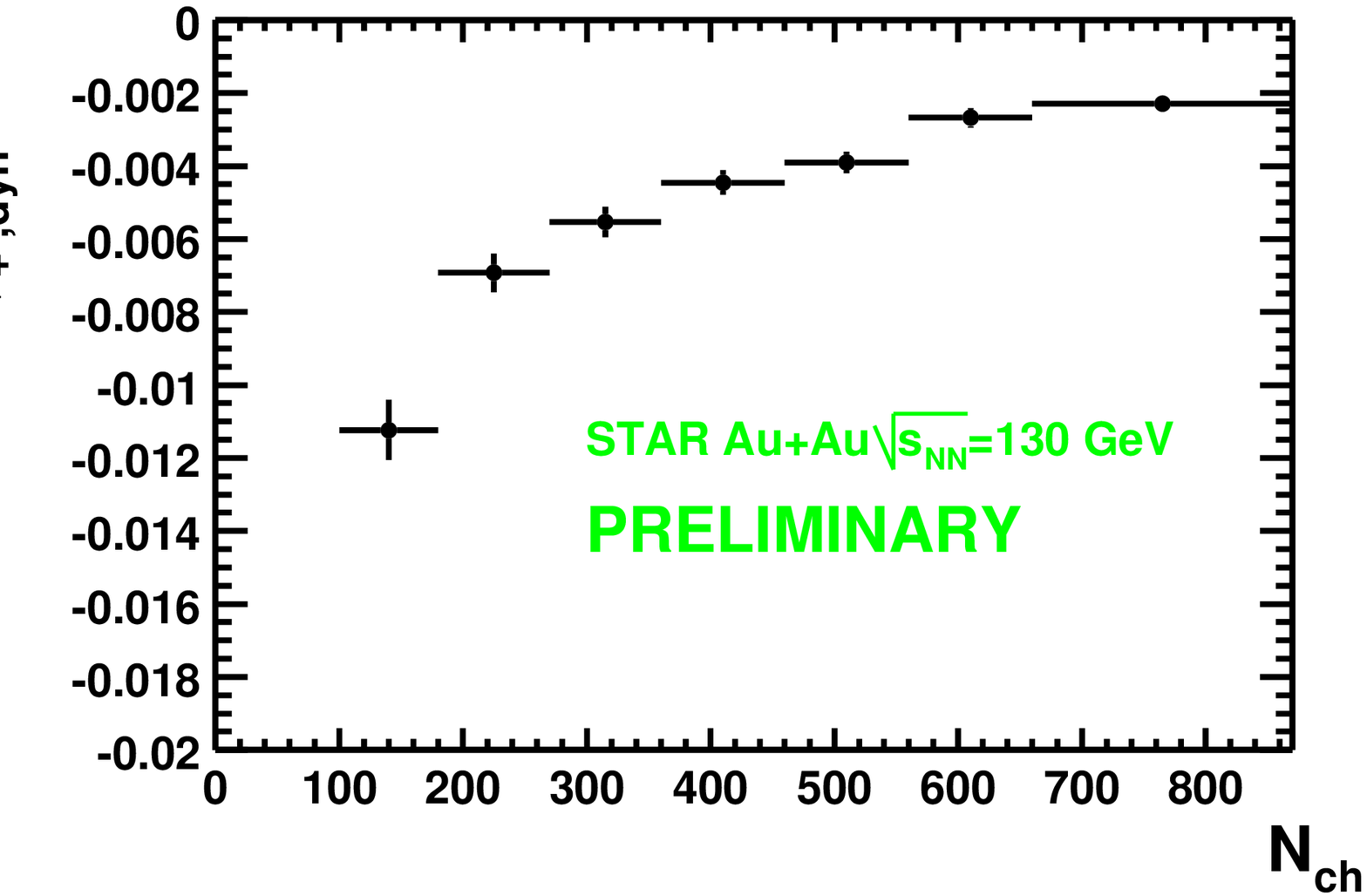}}
{\includegraphics[height=.25\textheight]{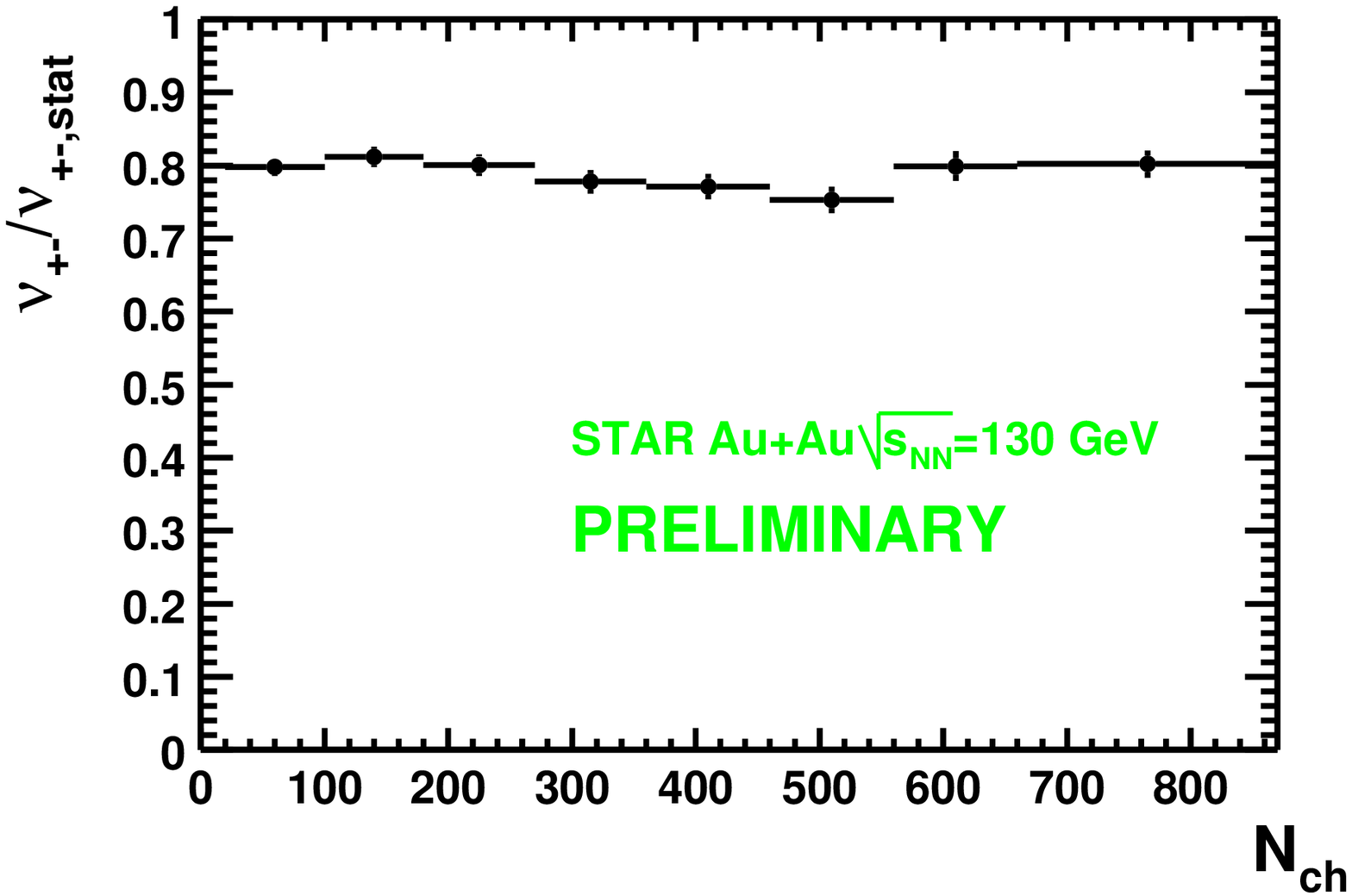}}
\caption{(left)  $\nu_{+-,dyn}$ 
as a function of centrality (event raw multiplicity), and (right)
$\nu_{+-}/\nu_{+-,stat} \approx 1+ (\la N_+\ra + \la N_- \ra) \nu_{+-,dyn} /4 $. }
\label{nu2}
\end{figure}

{\bf Multiplicity fluctuations.}
Fig.\ref{nu2} shows the dynamical fluctuations, $\nu_{+-,dyn}$, in
the pseudorapidity region $-0.5 < \eta <0.5$,
 as a function of the total multiplicity $N_{ch}$. 
The dynamical fluctuations are finite and negative: a clear 
indication of the correlation in the production of the positive 
and negative particles. 
We verified that, within statistical errors,  the same value 
is obtained when the pseudorapidity regions used to count positive an
negative tracks were separated by a $\Delta\eta=0.125$ gap, which
should eliminate problems related to track splitting and 
resolution of two close tracks.
We observe less than 10\% decrease in the strength of the correlation
when varying the pseudo-rapidity range from $|\eta|<0.125$ to 
$|\eta|<0.75$. 

The absolute strength of the dynamical fluctuations 
decreases monotonically with increasing collision centrality.
In order to check the deviation from the inverse proportionality to
multiplicity, in Fig.\ref{nu2}b we present the ratio 
of the observed fluctuations to the statistical 
fluctuations\footnote{ 
Note, that this ratio depends on the size of the acceptance used for the
measurement.}.
This  ratio is related to the quantities $\omega_Q$ and $D$ used in 
other works~\cite{Asakawa00,Jeon00,Gavin00} as $
\nu_{ab}/\nu_{ab,stat} \approx \omega_Q 
= \delta Q^2/N_{ch} = 0.25 D$.
The relationship is strictly exact if $\la N_+ \ra = \la N_- \ra$. 
The ratio has a value of about 0.8 nearly independent of the 
collision centrality in qualitative agreement with the expected value 
0.75 based on a thermalized resonance gas as 
calculated in~\cite{Jeon99} for particles produced
in a rapidity interval $\Delta y=1$. 
The measurement, however, deviates drastically from calculations
based on a rapid QGP freeze out ~\cite{Jeon00,Asakawa00}, 
which predict values of $\approx$0.2--0.4.

We consider these  results in the light of correlation functions 
measured in $p\bar{p}$ collisions at CERN~\cite{Foa75,UA5-88,Amendolia76}
with the use of Eq.\ref{nudyn}. 
We account for the unavailability of measurements at the same energy 
by interpolating the data obtained at lower and higher collision 
energies (parameterization from~\cite{UA5-86}). 
We also note that the correlation function for opposite charge 
particles, $R_{+-}$, is found to be approximately twice as strong 
as the same sign particles correlations, 
$R_{++}\approx R_{--}$~\cite{Foa75,Whitmore76}, 
and independent of the collision energy.
Using the single charge particle pseudorapidity 
density $\rho_1(\eta=0)\approx 2.05$  and 
$C_2(0,0)=\rho_2(\eta_1=0,\eta_2=0)-\rho_1(\eta_1=0) \rho_1(\eta_2=0)
\approx 2.8$  one finds charged-charged correlation function
$R_{cc} \approx 0.66$. 
Furthermore, assuming equal multiplicities of positively and 
negatively charged particles, one finds
$R_{cc}\approx (R_{++}+R_{--}+2R_{+-})/4 \approx 1.5 R_{++}$, 
which we use to estimate the correlation measured in this work as 
$R_{++}+R_{--}-2R_{+-}\approx -2 R_{++} \approx 4R_{cc}/3 \approx 0.88$. 
Under the assumption that the correlations are due to production in a 
finite number of sources (clusters), the correlation function 
 should be inversely 
proportional to the particle density.
In central Au+Au collisions, the charged particle density is about 550
compared to approximately 2.06 in $p\bar{p}$ collision. 
Such a dilution would give for the correlation function 
a value of $0.88 \cdot 2.06/550 \approx 0.003$, a number which is 
about 30\% larger than we observe (see Fig.\ref{nu2}a).

\begin{figure}[htb]
{\includegraphics[height=.25\textheight]{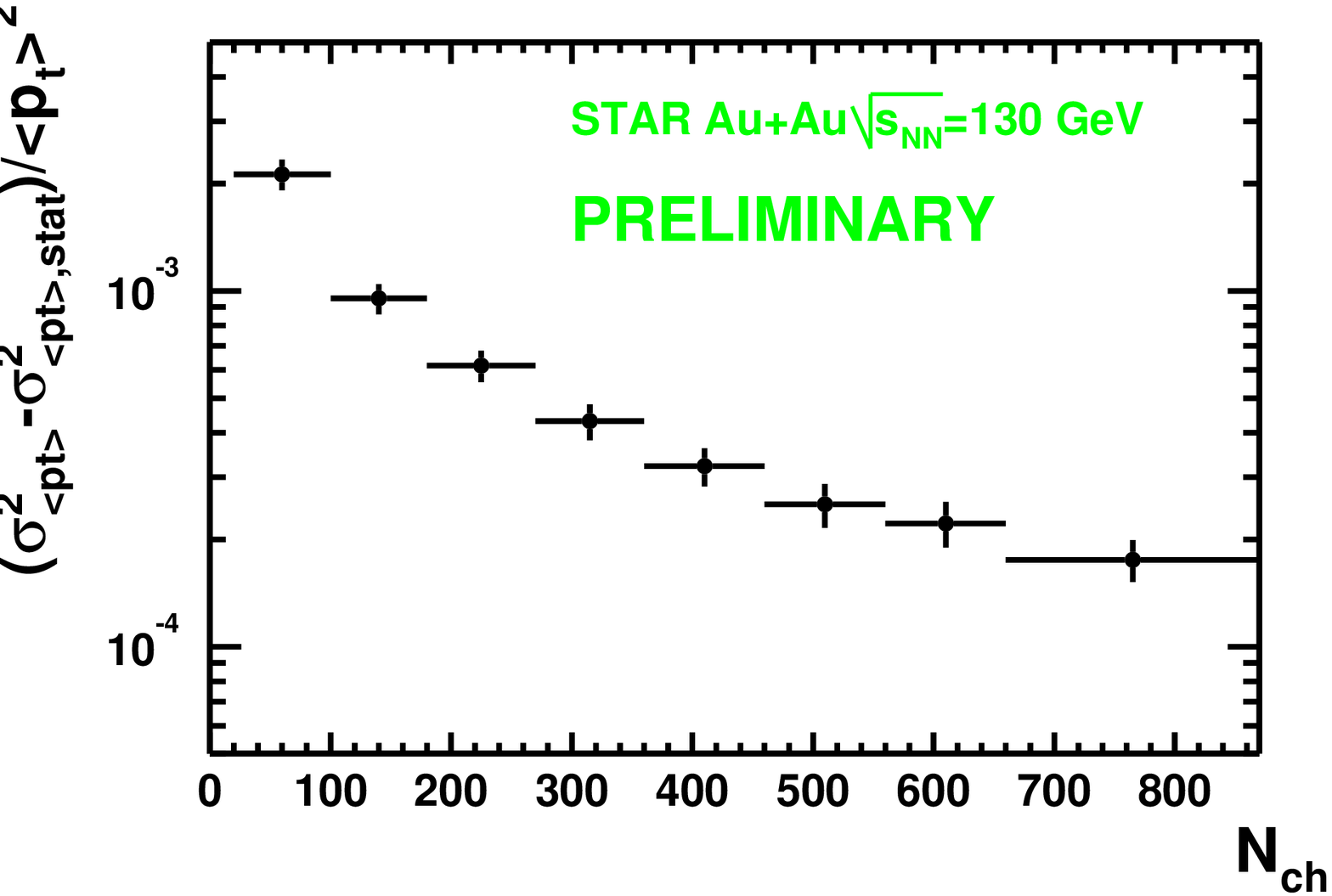}}
{\includegraphics[height=.25\textheight]{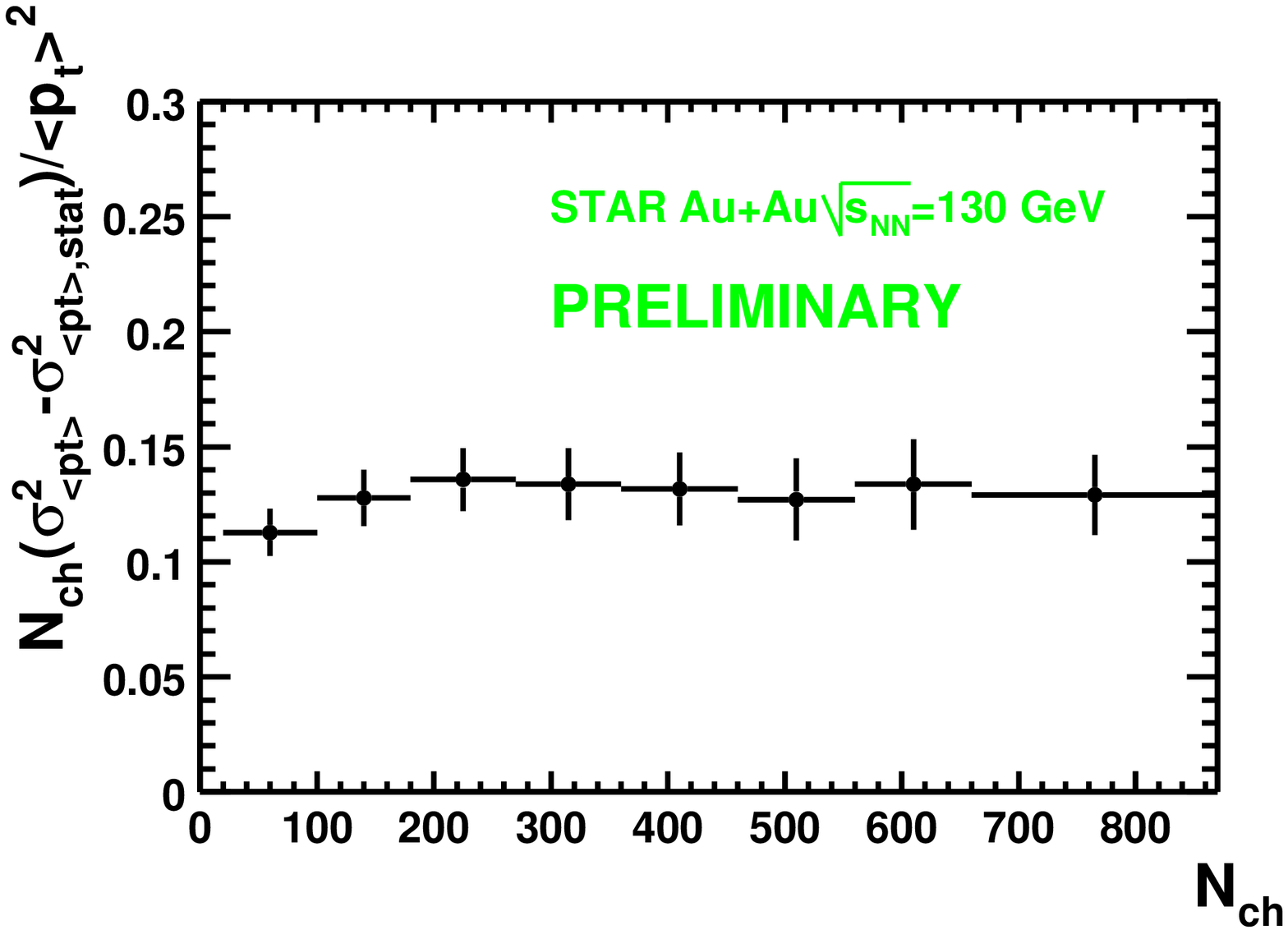}}
\caption{(left) $\sigmadyn^2/\mmpt^2$ as a function of centrality
(event raw multiplicity), and (right) the same scaled with multiplicity. }
\label{fig1}
\end{figure}

{\bf Mean $p_t$ fluctuations.} The results for the mean transverse
momentum fluctuation in the region $-0.75<\eta<0.75$ 
are presented in Fig.~\ref{fig1}.
Selecting specific pairs for the averaging in Eq.~\ref{sigmadyn} 
one can suppress
or enhance the contribution to $\mpt$ fluctuations of different origin.
We have also correlated:
(1) particles in pseudo-rapidity region $-0.75<\eta<-0.05$ 
with particles in $0.05<\eta<0.75$. The `gap' between
the two regions eliminates effects such as Bose-Einstein
correlations (HBT) or Coulomb final state interactions.  
(2) Positive particles with negative particles,
which is expected to enhance the contribution from resonances.
No significant difference has been observed.
The somewhat smaller value (less than 10\%) in the case 
of the correlation of particles
separated by the gap in pseudorapidity could be explained
by larger average distance in the pseudorapidity space
between the two particles in a pair. 
The relative fluctuation in mean transverse momentum was found to be 
$\sigma_{dynam}/ \langle \mpt \rangle \approx 1.2\pm0.2\% $.
The (relative) systematic error in our measurements of $\sigmadyn^2$ 
is about 5\%.  
It has been estimated by varying track selection cuts, 
primary vertex position within the TPC, and
comparing the Monte-Carlo results before and after processing through
the detector simulation.

Event-by-event dynamical fluctuations have also been analyzed by
several experiments using
the so called $\Phi_{p_t}$~\cite{phipt} measure
(the approximate relation to $\sigmadyn^2$ is taken from~\cite{vkr}): 
\begin{equation}
\Phi_{p_t} \equiv \sqrt{\la (\mpt- M\impt)^2 \ra/\la M \ra }-\sigmapt
\approx\frac{\sigmadyn^2 \langle M\rangle}{2\sigma_{p_t,inclusive}}.
\label{phiptdef}
\end{equation}
and close to it the difference factor $\Delta \sigma_{p_t}$~\cite{jeff}:
\be
\Delta \sigma_{p_t} \equiv (\sqrt{\la M \ra} \sigmapt
-\sigma_{p_t,inclusive}) \approx \Phi_{p_t}.
\ee
For the 6\% most central collisions STAR measures~\cite{jeff} 
$\Delta \sigma_{p_t} \approx 35~MeV$, which 
is consistent with results reported here for $\sigmadyn$ (one can
check this
with the help of an approximate expression given by Eq.~\ref{phiptdef} 
and using the other parameters presented in~\cite{jeff}).
In Pb+Pb and Pb+Au collisions 
at CERN SPS ($\sqrt{s_{NN}}=17$~GeV) $\Phi_{p_t}$ 
have been measured by NA49~\cite{na49plb} and 
NA45/CERES~\cite{ceres}. 
NA49 reported~\cite{na49plb} $\Phi_{p_t}=0.6 \pm 1$~MeV/c 
for the rapidity region $4.0<y<5.5$ and 5\% most central collisions.
The observed fluctuations are extremely small, but it would be incorrect to
compare these numbers with the STAR measurement, since NA49 results were
obtained in the forward rapidity region, 
while STAR measurements are at midrapidity.
The CERES collaboration has measured the fluctuations in the
central rapidity region. They report $\Phi_{p_t}=7.8 \pm 0.9$~MeV/c  
in central Pb+Au collisions~\cite{ceres}.
As seen from Eq.~\ref{phiptdef}, 
$\Phi_{p_t}$ is directly proportional to 
the number of reconstructed tracks used for its calculation 
(subject to acceptance cuts and tracking efficiency) which complicates
the comparison of the results from different experiments.
For a rough comparison one can take into account the fact that the CERES
multiplicity $\sim$130, $\langle \mpt\rangle \approx$420~GeV/c and
$\sigma_{pt,\; inclusive} \approx 0.270$~GeV/c. 
Then $\phisubpt \approx 8$~MeV corresponds 
to  $\sigmadyn/\langle \mpt\rangle \approx 1.4\%$. 

The dynamical part of mean transverse
momentum fluctuations has been measured at the ISR~\cite{isrpp}
in $pp$ collisions.
This was done by analyzing the multiplicity dependence of $\sigmampt$
under the assumption that in $pp$ collisions the dynamical part
in $\mpt$ fluctuations does not depend on multiplicity.
It was observed that $\sigmadyn/\langle \mpt\rangle \approx 12\%$.
Rescaling of this quantity with (the square root of) 
the ratio of multiplicity densities in
$pp$ and $Au+Au$ collisions yields the fluctuations in $Au+Au$:
$\sigmadyn/\langle \mpt\rangle \approx 0.8\%$, about 50\% less than
observed in this experiment.
 
To understand the role of (mini)jet production in $\mpt$ fluctuations
we have performed an analysis of central HIJING
events with hard processes switched ``on'' or ``off'' and  with the same cuts
as in the real experimental data. 
The jets increase the fluctuations  
from $\sigmadyn/\langle\mpt\rangle \approx 1.0\%$
to about 1.1\%.
Note that both numbers are very close to the experimentally observed value.
Contrary to HIJING, the RQMD event generator includes secondary 
particle rescatterings, but yields approximately the same values as
HIJING (within 20\%, limited
by our statistics of the RQMD events).

In {\bf summary}, we have reported measurements of the 
multiplicity and mean transverse momentum fluctuations in Au+Au
collisions at RHIC. While a clear correlation in particle production
has been measured, no qualitative difference from the lower energy
$AA$ collision experiments
as well as $pp$ collisions at similar energies has been observed.
The centrality dependence of the observed correlations is consistent
with being inversely proportional to total event multiplicity, the
dependence expected in particle production via clusters.


\doingARLO[\bibliographystyle{aipproc}]
          {\ifthenelse{\equal{\AIPcitestyleselect}{num}}
             {\bibliographystyle{arlonum}}
             {\bibliographystyle{arlobib}}
          }
\bibliography{flucta}

\end{document}